\documentclass[preprint,showpacs]{revtex4}
\usepackage{graphicx}
\usepackage{amsmath}

\begin{document}

\begin{titlepage}
\title{Nonequilibrium dynamics in an amorphous solid}
\author
{\it Sunil P. Singh and Shankar P. Das }

\affiliation {\it School of Physical Sciences, Jawaharlal Nehru
University, New Delhi 110067, India.}

\vspace*{1.5cm}

\begin{abstract}
The non-equilibrium dynamics of an amorphous solid is studied with a
soft-spin type model. We show that the aging behavior in the glassy
state follows a modified Kohlrausch-Williams-Watts (KWW) form
similar to that obtained in Phys. Rev. Lett. {\bf 95}, 055702 (2005)
from analysis of the dielectric loss data. The nature of the
fluctuation-dissipation theorem (FDT) violation is also studied in
the time as well as correlation windows.
\end{abstract}

\vspace*{.5cm} 

\maketitle
\end{titlepage}



In the glassy state, the liquid behaves like a frozen solid with the
motion of its constituent particles being localized around randomly
distributed sites. Analysis of the dynamics in this non equilibrium
glassy state reveals a variety of phenomena like aging and memory
effects \cite{ediger,struik}. Important progress in understanding
the non equilibrium dynamics of disordered systems has been made in
recent years from study of simple mean field spin glass models. In
the multi-spin interaction models, the non-linearities in the
Langevin dynamics give rise\cite{trk} to a ergodic-nonergodic
transition. The basic mechanism for this transition is very similar
to that present in the models for the dynamics of supercooled
liquids\cite{rmp}. The seminal work of Ref. \cite{cugli} dealt with
the problem of weak ergodicity breaking\cite{bouchaud} in a
spherical $p$-spin ($p>2$) interaction model\cite{soomers} over the
asymptotic time scales. Here the crossing over of the dynamics from
a regime of time translational invariance to that of aging behavior
was demonstrated analytically. Low temperature properties of glassy
systems, e.g., thermal conductivity and specific heat, have also
been studied with models \cite{kuhn} for the structural glass in
terms of a standard Hamiltonian involving spins. In the present
paper, we study a soft-spin type model, which is defined in terms of
the displacements of the particles around a corresponding set of
random lattice points. We show that the aging  behavior in the
nonequilibrium glassy state follows a modified KWW form similar to
that obtained in Ref. \cite{loidl} from analysis of the dielectric
loss data for several materials below the glass transition
temperature $T_g$.

We consider a model Hamiltonian, which has a translationally
invariant form in terms of the displacement variables $u_i$ around
an amorphous structure.

\begin{equation}
\label{haml-tr}
H= \sum_{p=2}^{\infty} \sum_{i\ne{j}}
J^{(p)}_{ij}(u_i-u_j)^p~~.
\end{equation}

\noindent For the amorphous solid, the interaction matrix
$J^{(p)}_{ij}$ is assumed to be random following a gaussian
probability distribution of zero mean and variance $J_p^2/N$. The
microscopic basis for such a model for an amorphous solid is
discussed further below. The time evolution of $u_i(t)$ is given by
the dissipative Langevin equation,
\begin{equation}
\label{langevin} \Gamma_{0}^{-1}\frac{\partial u_i}{\partial
t}~=~-\beta\frac{\delta H}{\delta u_i} - z(t)u_i + \xi_i(t).
\end{equation}
\noindent $\Gamma_0$ is the bare kinetic coefficient related to the
variance of the gaussian white noise $\xi_i$ through the
fluctuation-dissipation relation
$<\xi_i(t)\xi_j(t')>~=~2\beta^{-1}\Gamma_0\delta_{ij}\delta(t-t')$.
$z(t)$ is a Lagrange's multiplier used enforcing the constraint
$N^{-1}\sum_{i}<u_i^2(t)>=1$. This in the present context is
equivalent to having a constant Lindemann parameter at a fixed
temperature $T$. The simplest form of the nonlinear Langevin
equation (\ref{langevin}) is obtained by keeping in $H$ the
contributions from the $p=2$ and $3$ terms of the expansion
(\ref{haml-tr}). In the present work, we focus on the corresponding
nonlinear  model for the dynamics and refer this as the p23 model
from hereon.

We formulate a standard Martin-Siggia-Rose (MSR) \cite{msr} field
theory to compute the time correlation and response functions of the
displacement variable $u_i$. The field $\hat{u}_i$ conjugate to
$u_i$ is introduced in this regard to average over the gaussian
noise  $\xi_i$. The two time correlation and response functions are
respectively defined as : $C(t,t_\mathrm{w}) = N^{-1} \sum_{i=1}^{N}
\overline{\left<u_i(t) u_i(t_\mathrm{w})\right>}$, and
$R(t,t_\mathrm{w})= N^{-1} \sum_{i=1}^{N} \overline{\left<\hat
u_i(t) u_i(t_\mathrm{w})\right>}$, where the overbars stand for
averages over the random bonds $\{J^{(p)}_{ij}\}$'s and the angular
brackets represent mean over the gaussian white noise $\xi_i$'s. The
dynamics of the correlation and response functions are obtained from
the equations

\begin{eqnarray}
\label{cor-eq}
 \left [ {\partial}_t +z(t)\right ] C(t,t_\mathrm{w}) &=&
  \int_{0}^{t} ds \Sigma (t,s)C(s,t_\mathrm{w}) +
  \int_{0}^{t_\mathrm{w}} ds  ~\Xi^\prime (t,s)R(t_\mathrm{w},s) \\
\label{res-eq} \left [{\partial}_t +z(t) \right ] R(t,t_\mathrm{w})
&=& \delta(t-t_\mathrm{w}) + \int_{t_\mathrm{w}}^{t} ds \Sigma (t,s)
C(s,t_\mathrm{w}),
\end{eqnarray}

\noindent where  we denote $\Xi^\prime (t,t^\prime) =
2\delta(t-t^\prime)+\Xi (t,t^\prime)$. The kernels are obtained from
a perturbative summation as  $ \Xi(t,t^\prime)= \sum_p
a_{p}C^{p-1}(t,t^\prime),$ and $\Sigma(t,t^\prime)= \sum_p
(p-1)a_{p}C^{p-2}(t,t^\prime)R(t,t^\prime)$ in terms of a set of
coupling constants $\{a_p\}$, which depend on nonlinearities in the
dynamic equations. For the $p23$ model, we obtain up to one loop
order $a_2=2{(\beta J_2)}^2$ and $a_3=18{(\beta J_3)}^2$. The
necessary boundary conditions for $C$ and $R$ are respectively
chosen as\cite{cugli} : $R(t,t^-)=1$ and
$\partial_tC(t,t^{\pm})=\pm1$. The Lagrange's multiplier $z(t)$,
which ensures $C(t,t)=1$, is obtained as $z(t)=1 + \int_{0}^{t}
ds\left\{\Xi(t,s)R(s,t) + \Sigma(t,s)C(t,s) \right\}$.

The  analysis of the asymptotic dynamics of $C(t,t_\mathrm{w})$ for
both $t$ and $t_\mathrm{w}\rightarrow\infty$ is divided\cite{cugli}
into two main regimes. First, for $(t-t_\mathrm{w})/t\rightarrow 0$
the time translational invariance (TTI) holds. At this stage $C$ and
$R$ are related through the fluctuation dissipation theorem (FDT)
$R_\mathrm{I}(t)= -\Theta(t)\partial_{t}C_\mathrm{I}(t)$, where we
denote $C(t+t_\mathrm{w},t_\mathrm{w}){\equiv}C_\mathrm{I}(t)$ and
$R(t+t_\mathrm{w},t_\mathrm{w}){\equiv}R_\mathrm{I}(t)$. Second, for
$(t-t_\mathrm{w})/t \sim 1$, {\em i.e.}, for widely separated $t$
and $t_\mathrm{w} $ there is aging behavior. The correlation and
response functions, respectively denoted by $C_\mathrm{A}$ and
$R_\mathrm{A}$ in this case, are assumed to be functions of
$t_\mathrm{w}/t \equiv \lambda$ $(0<\lambda<1)$. We define
$C_A(t,t_\mathrm{w}) = q{\cal C}(\lambda)$ and
$R_A(t,t_\mathrm{w})=t^{-1}{\cal R}(\lambda)$. In the limit,
$\lambda\rightarrow{1}$, ${\cal C}(\lambda)\rightarrow{1}$ and
${\cal R}(\lambda) \ne{0}$. The solutions in the FDT and the aging
regimes agree if the long time limit of $C_\mathrm{I}(\tau)$ is $q$
termed as the non ergodicity parameter (NEP). In the FDT regime,
both the eqns. (\ref{cor-eq}) and (\ref{res-eq}) reduce to a single
equation

\begin{equation}
\label{tti} \left( {\partial }_t+1 \right)C_\mathrm{I}(t) +
\int_{0}^{t}ds \Xi_F(t-s)\partial_s C_\mathrm{I}(s) = z_\infty
[C_\mathrm{I}(t)-1 ].
\end{equation}

\noindent The kernel $\Xi_I [C_\mathrm{I}]$ reduce to $a_2
C_\mathrm{I}+ a_3C_\mathrm{I}^2$ in case of the p23 model. Except
for the linear term on the RHS, eqn. (\ref{tti}) is same  as the
basic dynamical equation in the self-consistent mode-coupling theory
of the structural glass. The latter represents the asymptotic
dynamics for the time correlation of the equilibrium density
fluctuations in a supercooled liquid. However, in the present case
the nontrivial renormalization contribution to the transport
coefficient comes from the dissipative nonlinearities in
(\ref{langevin}), while in the MCT for compressible liquids the
relevant nonlinearity is in the reversible pressure term. From the
$t\rightarrow\infty$ limit of eqn. (\ref{tti}) we obtain the
following relation

\begin{equation}
\label{cond2} \sum_p a_p(p-1)q^{p-2} +{(1-q)}^2 = 0~~.
\end{equation}

\noindent for the NEP $q$ in terms of the coupling constants $a_p$ .

In the aging regime, the FDT violation is denoted in terms of a
parameter $m$, which is defined through the relation
$R_\mathrm{A}(t)= -m\Theta(t)\partial_{t}C_\mathrm{A}(\tau)$ or
equivalently ${\cal R}(\lambda)=-mq(\partial/\partial\lambda) {\cal
C}(\lambda)$. We obtain analyzing eqns. (\ref{cor-eq}) and
(\ref{res-eq}) in the aging regime, the following equations  for $m$
and $q$,

\begin{equation}
\label{fdtv-par} m = (1-q) {{\sum_p a_{p}(p-2) q^{p-2} } \over
{\sum_p a_{p}q^{p-1}}}
\end{equation}

\noindent At the transition $m=1$. The critical coupling constants
$\{a_2^*,a_3^*\}$ for dynamic transition point of the p23 model is
obtained from the solution of eqns. (\ref{cond2}) and
(\ref{fdtv-par}) as $a^*_2=2/\lambda_0 - 1/\lambda_0^2$ and
$a^*_3=1/\lambda_0^2$, where $\lambda_0 =1-q$. The
ergodic-nonergodic transition line is given by
$a^*_2=2\sqrt{a^*_3}-a^*_3$. This is identical to the line of
dynamic transition in the $\phi_{12}$ model\cite{goetze,kim-mazenko}
of the mode coupling theory of structural glass transition. Along
the line of transition the parameter $\lambda_0$ changes from $0$ to
$0.5$ as the NEP changes from 1 to 0. In the ergodic phase, the NEP
$q=0$ and the FDT holds with $m=1$. Close to the transition line,
the relaxation behavior follows several regimes crossing over from
power law decay to a final stretched exponential form. The
corresponding stretching exponent $\beta_\alpha^E$ is
approximated\cite{kim-mazenko} with the empirical relation $\sum_p
a_p q p^{-1/{\beta_\alpha^E}}=1$.

\begin{figure}[t]
\centering
\includegraphics*[width=15cm]{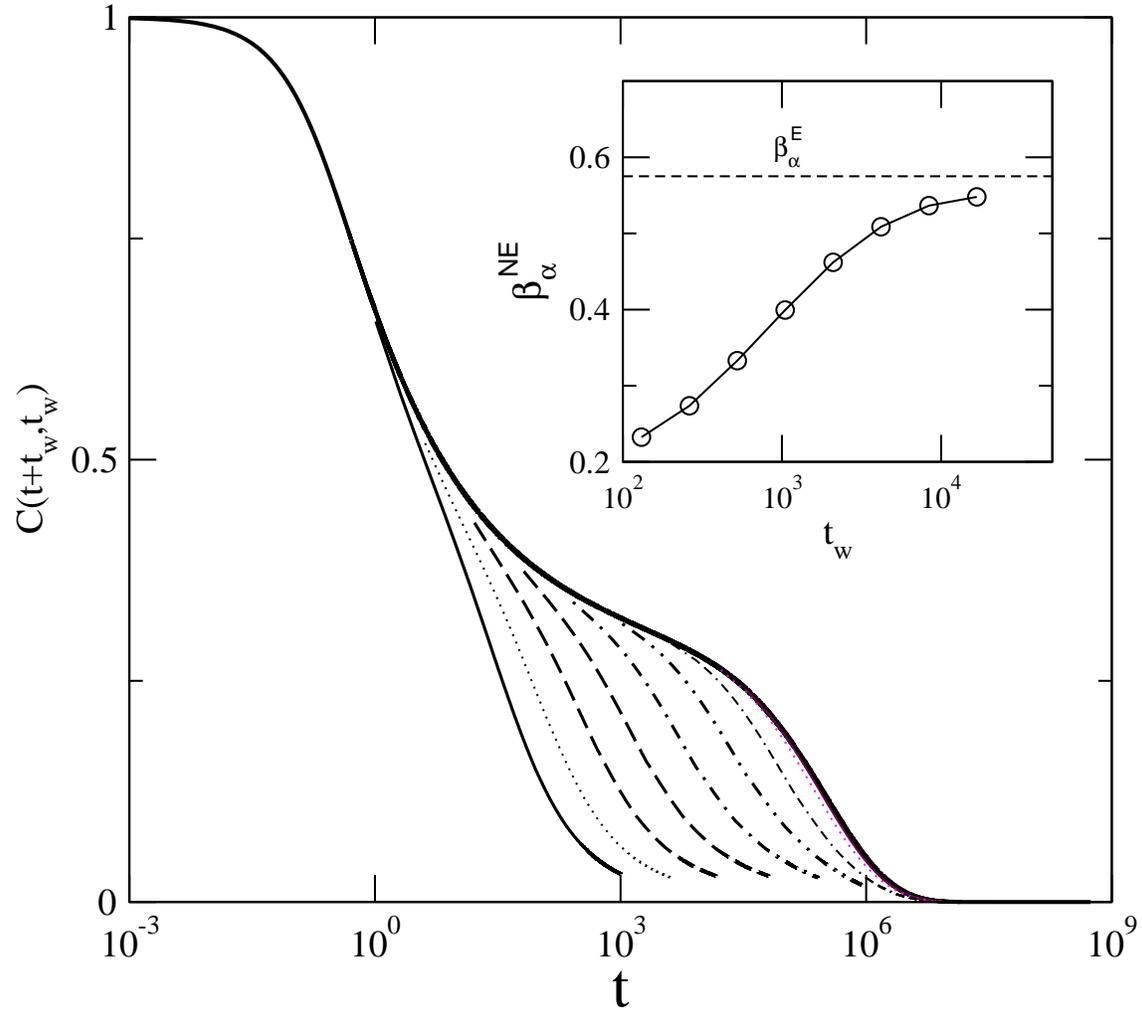}
\caption{The correlation $C(t+t_\mathrm{w},t_\mathrm{w})$ vs. $t$
for a set of waiting times $t_\mathrm{w}$'s in the ergodic phase,
$a_2=.82$ and $a_3=2.0$. Inset shows the exponent $\beta_{\alpha}$
for final stretched exponential relaxation w.r.t. $t_\mathrm{w}$.
Dashed line is the corresponding equilibrium value of stretching
exponent.}
 \label{fig1}
\end{figure}

To get a better understanding of the time scales associated with the
aging dynamics in the intermediate time regime and the corresponding
FDT violation, we solve the eqns. (\ref{cor-eq}) and (\ref{res-eq})
numerically. This requires integrating the equations (\ref{cor-eq})
and (\ref{res-eq}) for both $t$ and $t_\mathrm{w}$ extending over
several time decades. We use the adaptive integration technique
\cite{berthier}, which starts with smaller sized grids for
integration over shorter time scales of fast relaxation and
correspondingly increases the step size for longer time scales of
slow dynamics. In the ergodic state,  at long waiting times
$t_\mathrm{w}$ the correlation function approaches its equilibrium
value and time translational invariance is eventually reached. In
the approach to the equilibrium, the waiting time dependence of
$C(t+t_\mathrm{w},t_\mathrm{w})$ is displayed w.r.t. $t$ in fig.
\ref{fig1} for $a_2=0.82$ and $a_3=2.02$. In the final stage, the
decay follows the stretched exponential form
$\exp[-(t/\tau^{NE}_\alpha)^{\beta^{NE}_\alpha}]$ with
characteristic relaxation time $\tau^{NE}_\alpha$ and stretching
exponent $\beta^{NE}_\alpha$. The inset of fig. \ref{fig1} shows
$\beta^{NE}_\alpha$ corresponding to different waiting times
$t_\mathrm{w}$. At large $t_\mathrm{w}$, it approaches its
equilibrium value $\beta_\alpha^E$, which is determined in terms of
$a_2$ and $a_3$ using the empirical relation discussed above.

\begin{figure}[t]
\centering
\includegraphics*[width=15cm]{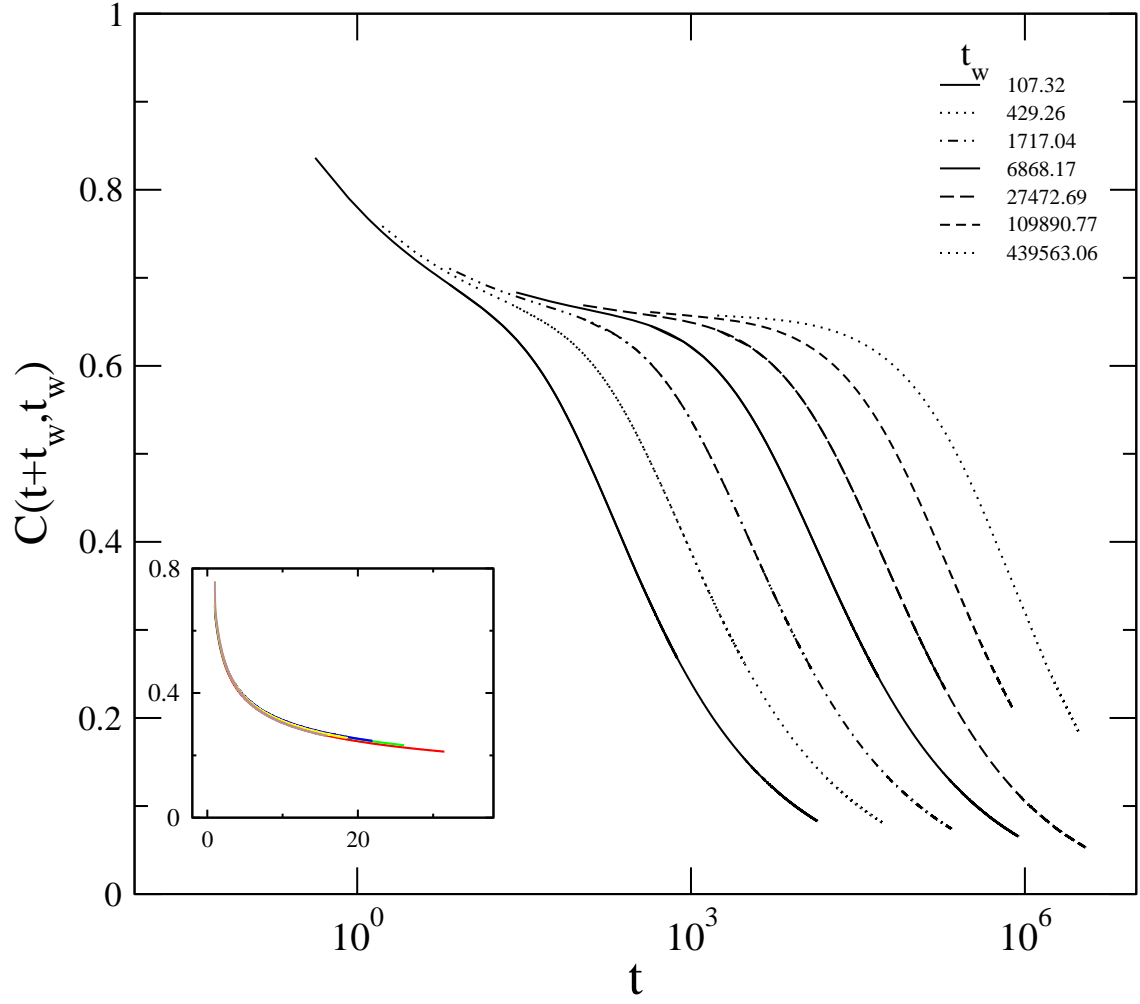}
\caption{The correlation $C(t+t_\mathrm{w},t_\mathrm{w})$ vs. $t$
for a set of waiting times $t_\mathrm{w}$'s in the nonergodic phase,
$a_2=0.5$ and $a_3=6.0$. Inset shows scaling of different
$C(t+t_\mathrm{w},t_\mathrm{w})$'s as a function of
$h(t+t_\mathrm{w})/h(t_\mathrm{w})$. Here
$h(t)\equiv\exp[t^{1-\kappa}/(1-\kappa)]$ with $\kappa=.96$.}
\label{fig2}
\end{figure}

In the non-ergodic state, the numerical solution of eqns.
(\ref{cor-eq}) and (\ref{res-eq}) displays both FDT and aging
behavior. In fig.\ref{fig2}, the time dependence of
$C(t+t_\mathrm{w},t_\mathrm{w})$ corresponding to $a_2=0.5$ and
$a_3=6.0$ deep in the glassy state are shown for different values of
$t_\mathrm{w}$. Initially the correlation decays from $1$ to $q$ and
at this stage time translational invariance holds. The dynamics is
strongly dependent on $t_\mathrm{w}$ at a later stage. The
corresponding correlation and response functions for large
$t_\mathrm{w}$  are scaled with the ansatz :
$C(t+t_\mathrm{w},t_\mathrm{w})=C [
h(t+t_\mathrm{w})/h(t_\mathrm{w})]$ where $h(t)$ is a monotonically
ascending function of $t$. The simplest possibility
$h(t)=t^{\gamma}$ is termed as the simple aging and implies
$C(t,t_\mathrm{w}) \equiv C(t/t_\mathrm{w})$. We adopt here the more
general form\cite{vincent,kim-latz}
$h(t)=\exp[t^{1-\kappa}/(1-\kappa)]$. The limit
$\kappa{\rightarrow}0$ implies time translational invariance while
$\kappa{\rightarrow}1$ represents simple aging. The case
$0<\kappa<1$ is termed as sub-aging. The dynamics almost conforms to
simple aging behavior as shown in the inset of fig.\ref{fig2} in
which different $t_\mathrm{w}$ data overlap on a single master curve
having $\kappa=0.96$. For every $t_\mathrm{w}$, the correlation
$C(t+t_\mathrm{w},t_\mathrm{w})$ decays to zero at sufficiently long
$t$. This is termed as weak ergodicity breaking in the aging regime.

\begin{figure}[t]
\centering
\includegraphics*[width=15cm]{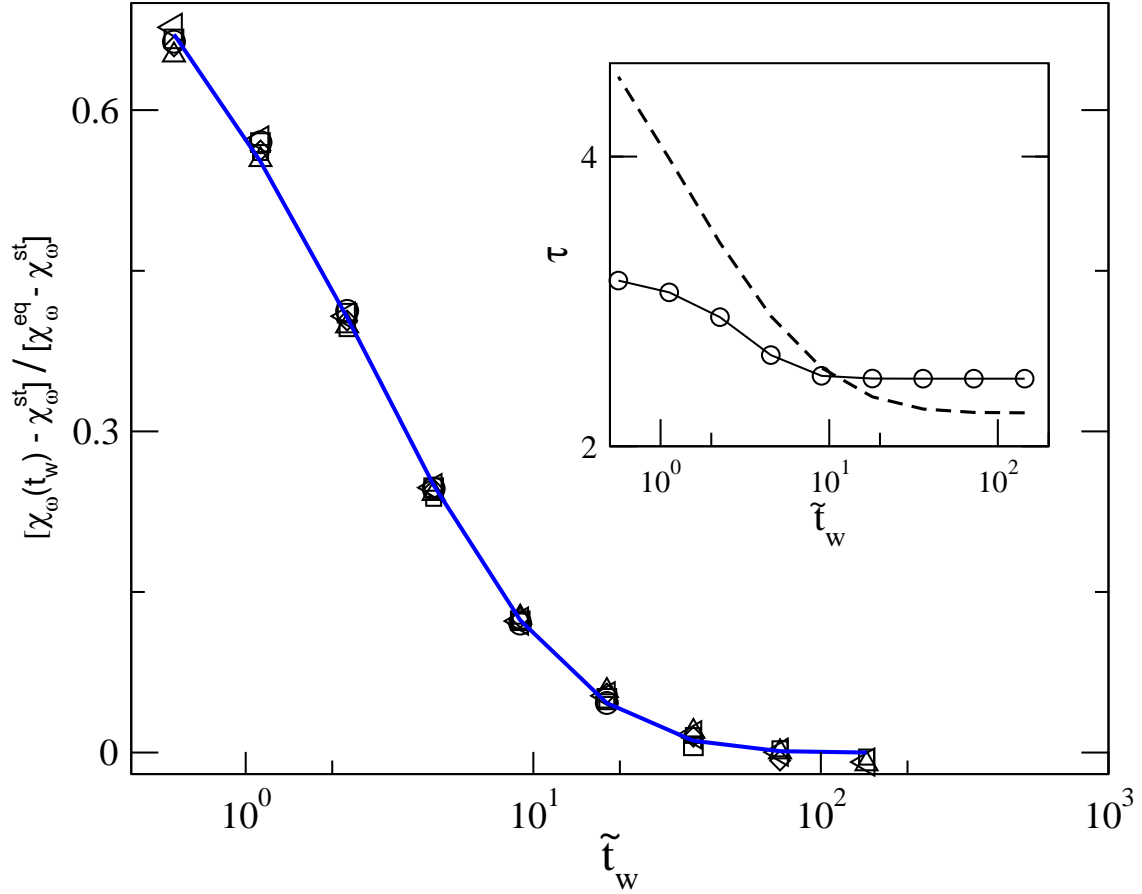}
\caption{Scaled function
$(\chi_{\omega}(\tilde{t}_\mathrm{w})-\chi_{\omega}^{st})/(\chi_{\omega}^{eq}-\chi_{\omega}^{st})$
vs. $\tilde t_\mathrm{w}$. Different symbols represent data for
different frequencies $\omega$. Solid line is a fit to modified KWW
form with $\tau(\tilde{t}_\mathrm{w})$ and $\beta=0.55$. Inset shows
the variation of the $\tau$ with $\tilde t_\mathrm{w}$, following
the scheme of (a) Ref.\citep{sengupta} (solid) and (b)
Ref.\citep{loidl} (dashed). } \label{fig3}
\end{figure}

\begin{figure}[t]
\centering
\includegraphics*[width=15cm]{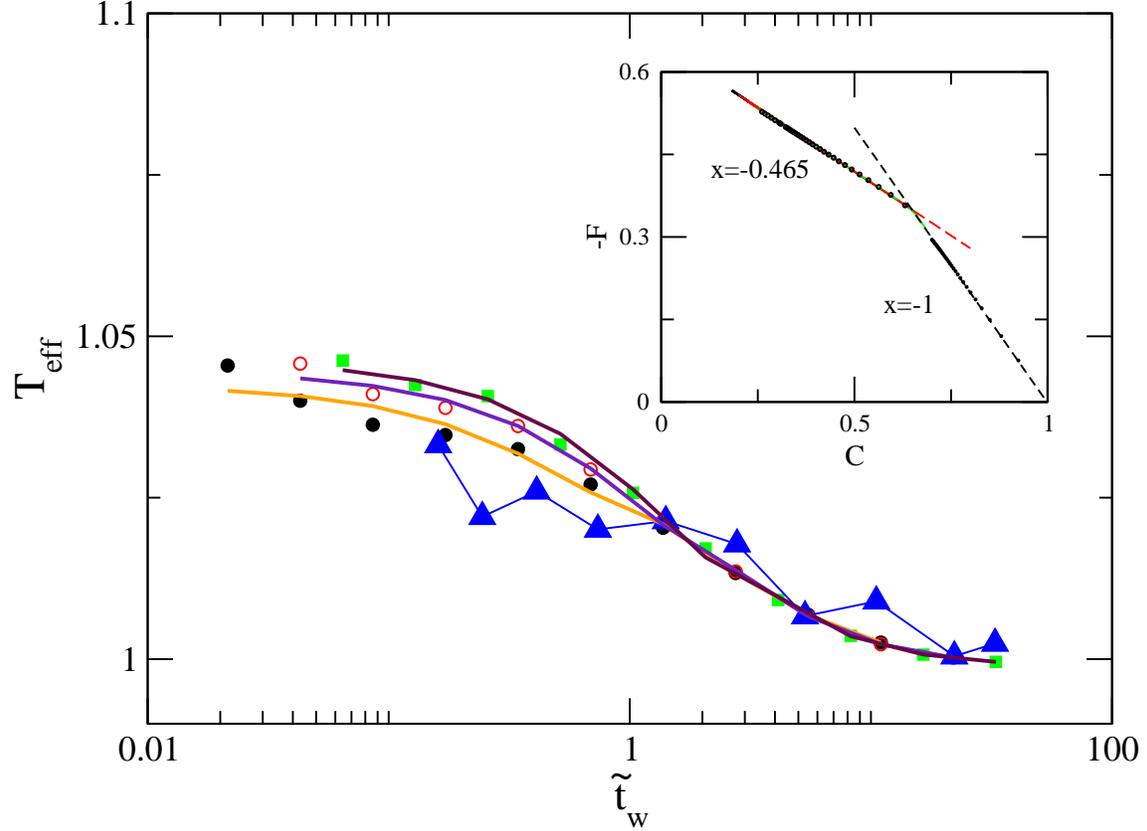}
\caption{Effective temperature $T_\mathrm{eff}$ (see text) vs.
waiting times $\tilde{t}_\mathrm{w}$ for $\omega=10^{-5}$(filled
circle),$3\times 10^{-4}$(open circle), and $5\times 10^{-4}$
(filled square) in units of . Solid line corresponds to best fit
curve of theoretical results. Experimental data  of Ref.
\citep{israloff} shown with filled triangles. Inset shows
$-F(t+t_\mathrm{w},t_\mathrm{w})$ vs.
$C(t+t_\mathrm{w}),t_\mathrm{w})$ corresponding to the data of fig.
\ref{fig2}. The slope in the FDT violation regime is $m=0.465$.}
\label{fig4}
\end{figure}

We now focus on the aging time dependence of the relaxation. For a
set of $t_\mathrm{w}$'s, the fourier transform of the correlation
function $C(t+t_\mathrm{w},t_\mathrm{w})$ with respect to $t$ is
obtained numerically. Since the correlation function in the aging
regime is approximately function of $t/t_\mathrm{w}$ ($\kappa=.96$),
the corresponding fourier transform $C(\omega,t_\mathrm{w})$ is a
function of ${\omega}t_\mathrm{w} \equiv \tilde{t}_\mathrm{w}$. For
comparison with experimental data, we define the response function
$\chi_\omega(\tilde{t}_\mathrm{w}) \equiv \omega
C(\omega,t_\mathrm{w})$. The waiting time ($\tilde{t}_\mathrm{w}$)
dependences of $\chi_\omega(\tilde{t}_\mathrm{w})$ for different
frequencies do not fit with a simple stretched exponential form
$\exp[-(t/\tau)^{\beta}]$ with constant $\tau$ over the whole time
range and a frequency independent $\beta$. The data is fitted  with
the modified KWW in a manner similar to that of Ref. \cite{loidl}.

\begin{equation}
\label{chi-dep}
 \chi_\omega (\tilde{t}_\mathrm{w}) = \left [
\chi^{\mathrm{st}}_\omega - \chi^{\mathrm{eq}}_\omega \right ]
\exp\left[-({\tilde{t}_\mathrm{w}}/\tau(\tilde{t}_\mathrm{w}))^{\beta}\right]
+ \chi^{\mathrm{eq}}_\omega
\end{equation}

\noindent where the subscripts "st" and "eq" respectively refer to
the limits $\tilde{t}_\mathrm{w} \rightarrow 0$ and $\infty$ for
$\chi_\omega$.  The aging time dependence of $\tau$ is chosen as

\begin{equation}
\label{tau} \tau(\tilde{t}_\mathrm{w}) = \left \{
\tau_{\mathrm{st}}- \tau_{\mathrm{fn}}\right \}
f(\tilde{t}_\mathrm{w}) + \tau_{\mathrm{fn}}~~,
\end{equation}

\noindent where $\tau_{\mathrm{st}}$ and $\tau_{\mathrm{fn}}$ are
fit parameters {\em independent} of frequency $\omega$. The
normalized function $f(s)$ is chosen to have limiting values $1$ and
$0$ for $s\rightarrow{0}$ and $\infty$ respectively. In particular,
we make the choice\cite{sengupta} $f(s)=
a_o/[1+\exp{\{s/\tau(s)\}}^\beta]$ where $a_0=2^\beta$ is a
normalization constant. Using this form of the
$\tau(\tilde{t}_\mathrm{w})$ we have fitted $\chi_\omega
(\tilde{t}_\mathrm{w})$ for all the different frequencies with a
single ( frequency independent ) stretching exponent $\beta$. In
fig. \ref{fig3}, a scaled plot of the different frequency data with
respect to $\tilde{t}_\mathrm{w}$ is displayed. The data sets for
all the frequencies merge on a single master curve with $\beta=0.55
$ and is shown as a solid line. For the dielectric loss data,
Lunkenheimer {\em et. al.} in Ref. \cite{loidl} use a somewhat
different fitting scheme with the $f(s)$ in eqn. (\ref{tau}) being a
stretched exponential function $\exp[-(s/\tau(s))^\beta]$. But these
authors adopt the parametrization of eqn. (\ref{tau}) not for the
time dependence of the relaxation time $\tau(\tilde{t}_\mathrm{w})$
in the modified KWW formula, but for the corresponding frequency
defined as $\nu
(\tilde{t}_\mathrm{w})=1/\{2\pi\tau(\tilde{t}_\mathrm{w})\}$.
Relaxation data when fitted with this scheme obtain the exponent
(also frequency independent) $\beta=0.53$. In the inset of
fig.\ref{fig4}, the $\tau$'s from both of the above described
fitting schemes are displayed. The stretching exponent values are
close although the relaxation time $\tau (\tilde{t}_\mathrm{w})$ are
in fact quite different in the two schemes.


We now consider the FDT violation in the nonequilibrium state. The
FDT is generalized in terms of a quantity $X(t,t^\prime)$ ( for
$t>t^\prime$ ) as $ k_BT R(t,t^\prime) = X(t,t^\prime)
{{\partial}C(t,t^\prime)}/{{\partial}t^\prime}$. In the limit
$t,t^\prime \rightarrow \infty$, it is assumed that $X(t,t^\prime)
\equiv x[C(t,t^\prime)]$ representing FDT violation in the
correlation windows rather than time windows. For convenience of
discussion, an integrated response function $F(t,t^\prime)$ is
defined

\begin{equation}
\label{new-G} F(t,t^\prime) \equiv \int_{t^\prime}^{t} ds R(t,s) =
\frac{1}{k_BT}\int_{C}^1 x(\bar{C})d\bar{C}~~.
\end{equation}

\noindent If the FDT holds, $x=-1$ and the above relation reduce to
$k_BT{F}(t)=C(t)-1$,  using  $C(t,t)=1$. An effective temperature
$T_\mathrm{eff}$ for the nonequilibrium state is defined in terms of
the ratio of the fourier transforms,
$k_BTF(\omega,{t}_\mathrm{w})/C(\omega,{t}_\mathrm{w})$. If the FDT
holds, $T_\mathrm{eff}=1$. Using the relation (\ref{fdtv-par}), we
obtain that the choice $a_2=0.5$ and $a_3=3.0$ in the p23 model
makes $T_\mathrm{eff}$ close to the experimental result of Ref.
\cite{israloff}. More importantly,  the time scale of
$\tilde{t}_\mathrm{w}$ over which the cross over from the FDT to the
aging regime occurs according to the present model is comparable
with experimental observations as shown in fig. \ref{fig4}. We also
display in the inset of this figure, the FDT violation corresponding
to the case of fig. \ref{fig2} as seen from the correlation windows.
This is similar to results\cite{barrat} from molecular dynamics
simulations of the binary Lennard-Jones mixtures.

The present model for an amorphous solid can be justified from a
semi-microscopic basis.  The potential energy is expressed as a Born
von Karman type expansion of the coordinates $\{r_i\}$ of the $N$
particles,

 \begin{equation}
 \label{pot-en} H= \sum_{ij}^\prime J^{(2)}_{ij}{u_i}u_j+
 \sum_{ijk}^\prime J^{(3)}_{ijk}{u_i}{u_j}{u_k}+..+ G(u_i),
 \end{equation}

\noindent where  $r_i = r^0_i + u_i$. The primes in the summations
in the RHS indicate that the terms having all the corresponding
 running indices $i,j,k$ etc. being same are absent. In case of the
 amorphous solid, $\{ r_i^0\}$ constitute a random structure
 corresponding to a local minimum of potential energy and $u_i$ is
 the displacement of the $i$-th particle from its parent site. The
 expansion in terms of $u_i$'s is valid over the time scale of the
 structural relaxation. The single site potential $G(u_i)$
 \cite{kuhn} in the RHS of (\ref{pot-en}) is being included to
 stabilize the system. We will approximate $U=\sum_{i<j} \phi_{ij}$
 as a sum of two body potentials and write the potential energy in
 the translationally invariant form given by eqn. (\ref{haml-tr}) by
  assuming
  $J^{(3)}_{ijk}=J^{(3)}_{ij}\delta_{jk}+J^{(3)}_{jk}\delta_{ki}+J^{(3)}_{ki}\delta_{ij}$ and $J^{(p)}_{ji}=(-1)^p J^{(p)}_{ij}$ etc.
 For reaching the expression (\ref{haml-tr}) the coefficients of the
 single site term $G(u_i)=\sum_i \{ w_{2i} u_i^2 + w_{3i} u_i^3 +
 ...\}$ are chosen as : $w_{2i}=-\sum^\prime_{j} J^{(2)}_{ij}$,
 $w_{3i}=-\sum^\prime_{j,k} J^{(3)}_{ijk}$, etc. The semi-microscopic
 interpretation described above is useful in linking the model with
 thermodynamic parameters\cite{sps-spd}. This will test further the 
 possibility of using the mode coupling approach to study the complex
 dynamics of the non-equilibrium state of an amorphous solid. CSIR,
 India is acknowledged for financial support.

\end{document}